\documentclass[final,1p,times]{elsarticle}
\usepackage{listings}
\usepackage{graphicx}
\usepackage{tikz}
\usepackage{amsmath}
\usepackage{amssymb}
\usepackage{multirow}
\usepackage{xcolor}
\usepackage{soul}
\usepackage{url}
\newcounter{bla}

\journal{Computer Physics Communications}

\begin{document}

\begin{frontmatter}



%---need a cool name
\title{{\tt QuDPy}: A Python-based Tool For Computing Ultrafast Non-linear Optical Responses}

\author[a,b1]{S. A. Shah\corref{author1}}
\author[a]{Hao Li\corref{author2}}
\author[a,b]{Eric R. Bittner\corref{author3}}
\author[d,c,e]{Carlos Silva\corref{author4}}
\author[b]{Andrei Piryatinski \corref{author5}}

\cortext[author3] {Corresponding author.\textit{E-mail address:} bittner@uh.edu}
\address[a]{Departments of Chemistry and Physics, University of Houston, Houston, Texas 77204, United~States}
\address[b]{Los Alamos National Lab, Los Alamos, New Mexico 87545, United States}
\address[b1]{CNLS and T-4, Los Alamos National Lab, Los Alamos, New Mexico 87545, United States}
\address[c]{School of Chemistry and Biochemistry, Georgia Institute of Technology, Atlanta, GA 30332, United~States}
\address[d]{School of Physics, Georgia Institute of Technology, Atlanta, GA 30332, United~States}
\address[e]{School of Materials Science and Engineering, Georgia Institute of Technology, Atlanta, GA 30332, United~States}
\begin{abstract}

Nonlinear Optical Spectroscopy is a well-developed field with theoretical and experimental advances that have aided multiple fields including chemistry, biology and physics. However, accurate quantum dynamical simulations based on model Hamiltonians are needed to interpret the corresponding multi-dimensional spectral signals properly. In this article, we present the initial release of our code, QuDPy (quantum dynamics in python) which addresses the need for a robust numerical platform for performing quantum dynamics simulations based on model systems, including open quantum systems. An important feature of our approach is that one can specify various high-order optical response pathways in the form of double-sided Feynman diagrams via a straightforward input syntax that specifies the time-ordering of ket-sided or bra-sided optical interactions acting upon the time-evolving density matrix of the system. We use the quantum dynamics capabilities of QuTip for simulating the spectral response of complex systems to compute essentially any $n^{th}$-order optical response of the model system. We provide a series of example calculations to illustrate the utility of our approach.

\end{abstract}


\end{frontmatter}

% \linenumbers

% All CPiP articles must contain the following
% PROGRAM SUMMARY.

{\bf PROGRAM SUMMARY}
  %Delete as appropriate.

\begin{small}
\noindent
{\em Program Title:}  {\tt QuDPy}                                       \\
{\em CPC Library link to program files:} (to be added by Technical Editor) \\
{\em Developer's repository link:}{  https://github.com/sa-shah/QuDPy} \\
% this will be a GitHub link
%
{\em Code Ocean capsule:} (to be added by Technical Editor)\\
{\em Licensing provisions:} 
MIT License (MIT License Copyright (c) 2022 Syed A shah)
\\
{\em Programming language:}                                 Python (v3.7)  \\
{\em Computer:} Any architecture with  Python (v3.7)  
\\
{\em Operating system:} Linux, MacOS
\\
{\em Supplementary material:}      Available as Google Colab Files. \\
Example 1: https://tinyurl.com/y3j5jmmr  \\
Example 2: https://tinyurl.com/37vwntn5 \\

{\em Required (External) packages:}
\begin{itemize}
    \item QuTip (v.4.7) and dependencies i.e. Numpy, Matplotlib ( https://qutip.org/) 
    \item UFSS Automatic Diagram Generator (https://github.com/peterarose/ufss)
\end{itemize}
{\em Has the code been vectorised or parallelized?:} 
Yes, parallelized within QuTip (v4.7) using MPI. 
\\
{\em Nature of problem(approx. 50-250 words):}\\
Accurate quantum simulations of complex systems 
are required in order to understand and interpret
multi-dimensional ultrafast spectroscopic signals. 
This code provides an open-source/multi-platform 
method that facilitates 
the generation of higher-order non-linear optical 
responses for an arbitrary molecular or material system given a model input Hamiltonian and bath model.
\\
  %Describe the nature of the problem here. \\
{\em Solution method(approx. 50-250 words):}\\
We use the double-sided Feynman diagram method
\cite{bib1,bib2}
to 
generate (symbolically) a set of response 
functions corresponding to the $n^{th}$ order 
non-linear response of the system to a series of 
laser pulses using the UFSS package\cite{bib4}
We then perform a series of 
accurate quantum 
dynamics calculations using the QuTip package\cite{bib3} to generate the numerical response
and spectra which correspond to specific experimental conditions. 

  %Describe the method solution here.
{\em Additional comments including restrictions and unusual features (approx. 50-250 words):}\\
None.
  %Provide any additional comments here.
   \\
%

%* Items marked with an asterisk are only required for new versions
%of programs previously published in the CPC Program Library.\\
\end{small}

\newpage
%% main text--this is the actual write up 
\section{Introduction}
\label{sec1}

Nonlinear, ultrafast spectroscopy 
has emerged as powerful experimental tool for 
probing coherent, light-induced processes 
in the condensed phase.
\cite{doi:10.1146/annurev-physchem-040513-103623,
SrimathKandada2022Homogeneous,
OGILVIE2009249,
doi:10.1146/annurev.physchem.51.1.691,
doi:10.1146/annurev.physchem.54.011002.103907,
doi:10.1021/jacs.9b10533} 
In this technique, 
one prepares a sequence of ultra-narrow (in time) optical 
pulses,  with specific phase matching conditions, that 
induce a macroscopic polarization response.  This polarization is directly linked to the 
time-evolution of the sample under investigation. 
By varying the time intervals between phase-matched
pulses one can develop a multi-dimensional map correlating
the absorption at one time with emission at some later time. By comparing the experimental signals against robust
theoretical models, one can gain tremendous insight into 
the inner quantum dynamics of a complex system. 

The field itself is mature and two textbook-level works are considered as the general introduction in this field.
\cite{Mukamel1995,hamm_zanni_2d_spectroscopy_2011}
  However, setting up and performing accurate quantum simulations of complex systems, along with their interaction with the
  thermal environment
 has long been the purview 
of theoretical groups.  As a result, the 
community currently
lacks a universal and portable code for performing
such simulations.  We present our
implementation of a general platform for
computing the ultrafast optical response for a 
generalized system that can be described in terms
of a model Hamiltonian in contact with 
a dissipative environment. Our platform can simulate a wide range of molecular and condensed matter systems with finite-dimensional model Hamiltonians. It offers simulation capabilities across a broad spectral range (from NMR to UV-Vis), encompassing various transitions such as electronic, vibrational, rotational, and more. Our code, written in Python 3, leverages the open-source QuTip quantum simulation package.\cite{QuTip2}  It provides the user with a straightforward way to automatically generate and compute non-linear responses, as specified by a double-sided Feynman diagram given as input. The double-sided diagrams are created using the Automatic Feynman Diagram Generator within the UFSS package, tailored to the desired phase-matching or phase-cycling condition.\cite{UFSS1, UFSS2}

Our platform currently simulates non-linear responses in the impulsive regime, applicable to a significant number of ultra-fast optics experiments. However, this simulation does not include electric field polarization. Future releases will enhance the capabilities by covering both the non-impulsive regime and resolving electric field polarization.

It is worth pointing out the other notable efforts on computational fronts in ultrafast optical spectroscopy. The Spectron/iSpectron by Mukamel et. al. provides the spectral calculations via two methods. Firstly, though the average amplitudes of Liouville pathways and cumulant expansion of Gaussian fluctuations in reference eigen-energies; and secondly with the use of the molecular configuration specific Green's function expressions.\cite{Spectron} On the other hand, the Ultrafast Spectroscopy Suit (UFSS) by Ross and Kirch obtains the nonlinear spectroscopic signatures (for both closed and open quantum systems with finite-dimensional Hilbertspace) via perturbative expansion of wavefunction, and conversion of light-matter interaction with subsequent evolution into a convolution problem with appropriately designed operators. Their method takes into account the temporal profiles of the laser pulses. This results in reduced computational cost for simulation, especially in the case of temporal overlap between different laser pulses. Additionally, the suit provides automatic generation of double-sided Feynman diagrams that has been utilized in this work. \cite{UFSS1, UFSS2}  Briefly, our approach differentiates itself from these routines by computing the direct evolution of density matrix for the system with intermittent projection with light-matter interaction defining and guiding specific Liouville pathways.

We begin with a general overview of the theory 
of multi-dimensional spectroscopy, 
its implementation in our code, and 
describe its installation and required 
components. 
We then present a few model calculations showing how 
one can setup, compute, and interpret the results.
Our intention is that the code provides a robust and adaptable 
platform for both theory and experimental groups
working in this area. 
To maintain the brevity of the article, the complete code for these simulations is available in Examples 1 and 2 on the Google Colab platform and the input parameters for various functions and methods in the package are thoroughly explained in the Appendix.

\section{Theoretical 
background and 
description of the 
algorithm}

In any spectroscopic experiment, one 
ultimately is measuring a macroscopic polarization 
of the sample as induced by an incident 
applied electric field and one
can express the polarization in terms of powers 
of the electric field $\vec E$
\begin{align}
    \vec P = \epsilon_o\left(\chi^{(1)}\cdot \vec E
    + \chi^{(2)}\cdot \vec E\cdot \vec E
    + \chi^{(3)}\cdot \vec E\cdot \vec E\cdot \vec E
    +\cdots\right)
\end{align}
where $\chi^{(n)}$ are susceptibilities.
Since the electric field is a vector, the various 
susceptibilies are in fact tensors.
In isotropic media with inversion symmetry, 
the polarization must change
sign when the optical electric fields are reversed. Consequently, all the even-order terms vanish and the 3rd order term is the lowest-order non-linearity.

Under the dipole approximation in light-matter interaction, the observed polarizability is calculated by taking the expectation value of the quantum dipole operator:
\begin{align}\label{polSeries}
    P(t) = \text{Tr}(\mu\rho(t)) = \langle \mu \rho(t)\rangle
\end{align}
where $\rho(t)$ is the quantum density matrix for the
system which has evolved under the influence of the 
external optical fields. 
This, too, can be expanded in powers of the 
electric field 
as 
\begin{align}
    P(t)  = \sum_{n=1}^{\infty}P^{(n)}(t)
\end{align}
Each of the terms in this series can be computed 
within the context of time-dependent 
perturbation theory by 
assuming that the applied fields 
interacts with the system at times
$\tau_1 < \tau_2 < \cdots \tau_n$. 
The resulting expression takes the form
of a series of nested time integrals 
and commutation operations, 
\begin{align}
  P^{(n)}(t)
 & = \left(-\frac{i}{\hbar}\right)^n
  \int_{-\infty}^t d\tau_n
   \int_{-\infty}^{\tau_n} d\tau_{n-1}
   \cdots
      \int_{-\infty}^{\tau_2} d\tau_{1}
      E(\tau_n) E(\tau_{n-1})\cdots  E(\tau_1)
      \nonumber 
      \\
    &\times  
      S^{(n)}(1,2\cdots,n) ,
\end{align}
where $S^{(n)}(1,2\cdots,n)$  is the  susceptibility
given as a series of interactions with 
the dipole operator $\hat\mu$ interspersed 
by propagation under the Liouville super-operator
representing the system dynamics, 
\begin{align}
    S^{(n)}(1,2\cdots,n) 
    = \langle 
   \hat \mu(n) e^{i{\cal L}t_n}[
   \hat\mu(n-1), \cdots
    e^{i{\cal L}t_{2}}
    [\hat \mu(2), e^{i{\cal L}t_{1}}[\hat\mu(1),\rho(0)]]]
     \rangle
\end{align}
Note that we have indexed 
each interaction with the light field by
the order in which it appears in the time sequence,
\begin{align}
    \tau_1 = 0 \nonumber \\
    t_1 = \tau_2 - \tau_1 \nonumber\\
    t_2 = \tau_3 - \tau_2 \nonumber\\
    \vdots\nonumber
    \\
    t_n = t-\tau_n\nonumber
\end{align}
and reflects the specific timing of the experimental 
laser pulses. 
This is defined for all positive times $t_n$.  
The
difficulty, thus far, is that one obtains 
$2^n$ terms since the dipole operator can 
act either on the left-side or right-side of the 
density matrix. However, each term is
paired with its complex conjugate, giving 
$2^{n-1}$ unique terms. 
Further, for an $n^{th}$ order non-linear response, the
electric field itself is composed of $n$ pulses,
\begin{align}
    E(t) = \sum_{i=1}^n E_i(t)(e^{-i\omega t+\vec k\cdot \vec r} + e^{+i\omega t - \vec k\cdot \vec r}),
\end{align}
such that the simplest 3rd order expression 
contains $6 \times 6 \times 6 \times 4 = 864$ 
possible terms.  However, 
most of these terms either vanish or are 
equivalent. 
Within the field of ultrafast spectroscopy, 
these terms are commonly expressed in a 
 diagrammatic
representation 
whereby propagation in time is represented as a vertical line and interactions via the density 
matrix are incoming or outgoing arrows.
\cite{Mukamel1995,hamm_zanni_2d_spectroscopy_2011}
The last interaction, which is not within the nested 
commutators, represents an outgoing field. The wavevector and frequency of this field comply with the conservation of momentum and energy from previous interactions, resulting in phase-matching and phase-cycling conditions; such that the final wavevector is the sum of wavevectors from all previous interactions and is able to distinguish rephasing and non-rephasing pathways.
Following the usual conventions in this field, one has a 
series of rules:
\begin{enumerate}
    \item The two vertical lines represent the time evolution of the ket and bra sides
    of the density matrix with time running from the bottom to the top.
    \item Interactions with the light field 
    are represented as arrows entering or
    leaving the diagram at specified times.
    Since the last interaction occurs outside the 
    nested commutator, it is different and is represented always as an outgoing (dashed) arrow.
    \item Each diagram has a sign $(-1)^n$ where
    $n$ is the number of interactions on the right
    side of the diagram. This is due to the fact that
    each right-hand side interaction carries a minus
    from the commutator bracket.  The last interaction 
    originates outside the bracket and does not contribute a sign. 
    \item An arrow pointing to the right represents
    an electric field with $e^{-i\omega t + ikr}$
    while one pointing to the left represents an
    electric field with $e^{+i\omega t - ikr}$.
    The emitted light (last outgoing arrow) has a frequency
    and wavevector given by the sum of the input frequencies and wavevectors. 
    \item  Arrows pointing towards the system
    correspond to excitations of the system 
    while those pointing away are de-excitations. 
    \item The final trace operation requires that the
    system be in a population state after the last 
    interaction. The population state in this context is of the form $|n\rangle \langle n|$ as opposed to a coherence, represented by $|n\rangle \langle m|$.
    \item By convention, we will show only diagrams with the last interaction emitting from
    the ket (i.e. the left side). 
\end{enumerate}

\begin{figure}
    \centering
 \includegraphics[width=\columnwidth]{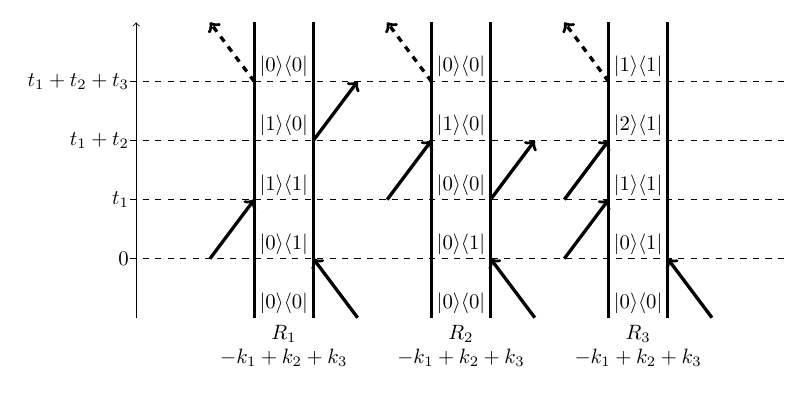}
  \includegraphics[width=\columnwidth]{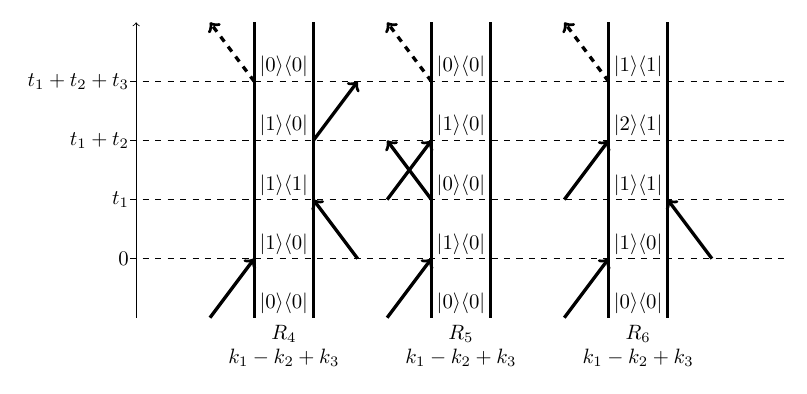}
    \caption{(Top Row) Rephasing double-sided diagrams for 3rd order response.
    In a rephasing experiment, the 
    outgoing signal emerges at $k = -k_1 + k_2 + k_3$.
    (Bottom Row) 
    Non-rephasing double-sided diagrams for 3rd order response.
    In a non-rephasing experiment, the
    outgoing signal emerges at
     $k = +k_1 - k_2 + k_3$. The numbering and notation of these diagrams follow the convention in Ref. \cite{hamm_zanni_2d_spectroscopy_2011}
    }
    \label{fig:diag}
\end{figure}

The picture that emerges is that at time $\tau_1 = 0$,
the system interacts (from the left or right) with  
the impinging laser field, which projects 
a coherence 
within the system. For example, for a 
two state system starting its ground state with 
$\rho(-\infty) = |0\rangle \langle 0|$
we would have that
$\rho(0^+) = 
(|1\rangle\langle 0| +|0\rangle\langle 1|) $.
Each term now evolves under the Liouvillian for time $t_1$
such that
\begin{align}
    \rho(t_1) = e^{i{\cal L} t_1}\rho(0^+)
\end{align}
At time $t_1$, one has an interaction followed
by a time evolution $t_i$, until the last projection
at time $t_n$ when the final signal is observed.
Since the projections can be on the right 
or on the left side of the density matrix, 
a given experiment can be thought of as a 
sum over all left and right side interactions. 
The diagram  rules lead to four distinct 3rd order diagrams for a 2-level system and six for a 3-level system (for rephasing and non-rephasing signals, or, equivalently, photon echo and virtual echo spectra). 
The first four are shown on the left in Fig.~\ref{fig:diag}, whereas the additional two are shown on the right.

\begin{table}
\begin{tabular}{ |p{3cm}||p{3cm}|p{5cm}| }
 \hline
 \multicolumn{3}{|c|}{Interaction Diagrams} \\
 \hline
 Response Type & Action & Input Syntax\\
 \hline
 $R_1(t_1,t_2,t_3)$ & $[B^+,K^+,B,K]$ & {\tt ((`Bu',0),(`Ku',1),(`Bd',2))} \\
$ R_2(t_1,t_2,t_3)$ & $[B^+,B,K^+,K]$ & {\tt ((`Bu',0),(`Bd',1),(`Ku',2)) }\\
$ R_3(t_1,t_2,t_3)$ & $[B^+,K^+,K^+,K]$ & {\tt((`Bu',0),(`Ku',1),(`Ku',2))}\\
$ R_4(t_1,t_2,t_3)$ & $[K^+,B^+,B,K]$ & {\tt((`Ku',0),(`Bu',1),(`Bd',2))}\\
$ R_5(t_1,t_2,t_3)$ & $[K^+,K,K^+,K]$ & {\tt((`Ku',0),(`Kd',1),(`Ku',2))}\\
$ R_6(t_1,t_2,t_3)$ & $[K^+,B^+,K^+,K]$ & {\tt((`Ku',0),(`Bu',1),(`Ku',2))}\\
 \hline
\end{tabular}
\caption{Summary of response types, operations, and syntax for a 3rd order coherent process corresponding to the 
diagrams shown in Fig.~\ref{fig:diag}.}
\label{table:1}
\end{table}

For computational purposes, any diagram can be compactly represented 
as a time-ordered list of left or right-side operations such as 
where $K^+$ and $K$ denote
an excitation and a de-excitation on the left (i.e. on the ket), respectively. Similarly, $B^+$ and $B$ denote
a de-excitation and an excitation on the right (i.e. on the bra), respectively. 
Following the convention introduced by Automatic Feynman Diagram Generator, in QuDPy, these interactions are encoded by `Ku' and `Kd' for $K^+$ and $K$; similarly, $B^+$ and $B$ are represented by `Bu' and `Bd', respectively.\cite{UFSS1, UFSS2}
Table ~\ref{table:1} gives a summary
of 6 possible response functions for 
a coherent 3rd order process with their 
equivalent actions and corresponding QuDPy syntax.
Notice that the last interaction is not indicated in QuDPy syntax
since it always corresponds to a ket-side de-excitation operation by convention.

\begin{figure}[t]
    \centering
 \includegraphics[scale=0.8]{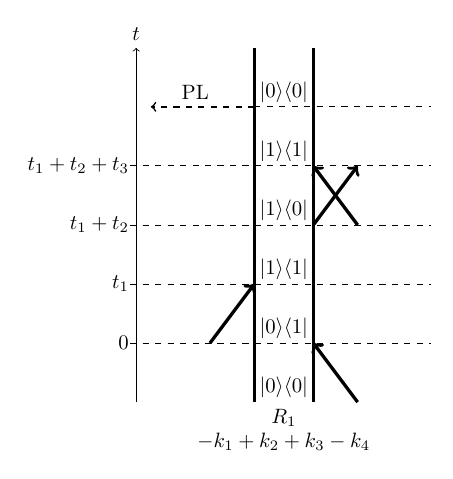}
    \caption{An example of a 4-th order rephasing diagram for an excited-state emission process specified by
    {\tt  ((`Bu',0),(`Ku',1),(`Bd',2),(`Bu',3))}.
    }
    \label{fig:4th}
\end{figure}

Note, although only odd-order terms contribute to the polarization in isotropic media, one can also consider interactions which are even-order
with respect to the density matrix by measuring the photo-luminescence (PL) rather than the polarization.
Here, we assume that the final signal is proportional to the relatively slow, spontaneous emission of light from an excited population state, (in contrast to the prompt, stimulated emission from a coherence state in odd-order spectroscopes).\cite{doi:10.1063/1.2800560} 
In this scenario, one can use a 4-pulse sequence 
as sketched in Fig.~\ref{fig:4th}.
in which the out-going PL 
reflects the excited state population \textit{vis}.
\begin{align}
    PL_n^{(r)}(t) \propto \langle n|
    \rho^{(r)}(t)|n \rangle
\end{align}
where $\rho^{(r)(t)}$ is the
$r^{th}$ order term in the perturbation expansion of the 
density matrix.   
This is one of a number of 
possible interaction pathways 
and can be input as 
\begin{align}
    R_1(0,1,2,3;t)={\tt  (('Bu',0),('Ku',1),('Bd',2),('Bu',3))}
    \end{align}
along with 4 times, $\{t_1,t_2,t_3,t\}$,
in which the $t_2$ corresponds to the 
population waiting time. 
The last time $t$ corresponds to the observed PL.
Additionally, in a comparable experiment, these terms can determine the photo-current generation, with the observed photocurrent being proportional to the population of the final excited state.
\cite{Karki:2014aa,Bakulin:2016aa,Vella:2016aa,LI2016281}

Finally, once response elements (R's) are obtained, the susceptibility can be computed by summation of constitutive responses.
Typically, one works in a state space 
in which the system Hamiltonian is purely
diagonal and the interaction with the 
external laser field is treated as strictly
non-diagonal perturbation. As an example, 
if we take the ``system'' to be a harmonic oscillator with $H_o = \hbar\omega (a^\dagger a + 1/2)$ 
and the light-matter interaction to be mediated by the dipole operator: $\hat\mu = \mu (a+a^\dagger)$,
one can immediately write down the various response functions in Table \ref{table:1}. 
For example, 
\begin{align}
    R_1(t_1,t_2,t_3) =(-1)^2\mu^4\langle a(t_3) a^\dagger(t_1)\rho(-\infty)a(t_0)a^\dagger(t_2)\rangle
\end{align}
in which $a(t)$ is the harmonic oscillator 
destruction operator in the interaction representation, i.e $a(t) = e^{i\omega t}a$.
However, a generalized physical system is never this simple since the dipole operator 
can create transitions between multiple 
quantum states. Moreover, interaction with 
the environment can induce loss of phase coherence and non-radiative relaxation processes that need to be accounted for 
in a generalized computational package. 

With this in mind, 
our implementation uses the QuTiP 4.x package
to handle all of the setup and propagation of the
system variables.\cite{QuTip2}  
Our approach assumes that one can 
write down the system Hamiltonian 
and external coupling 
in terms of a standard set of quantum 
operators, such as the spin-matrices or harmonic oscillator operators.  This includes 
systems with time-dependent Hamiltonians, systems interacting with complex bath degrees of freedom, 
cavity QED, polaritons, and a wide variety 
of open-quantum systems. 
The current version 
of QuTip includes both Lindblad and Redfield equation
integrators which are continuously improved upon to optimize 
both numerical efficiency and memory overhead. For conciseness, we omit details on the methods and capabilities for defining system Hamiltonians and system-bath interactions, as they are thoroughly discussed in the documentation of the QuTip package.\cite{qutipDoc}

Our implementation  
interrupts the QuTip {\tt mesolve()} routine at 
the specified intervals, performs the required 
projection, and then allows the propagation to continue. 
Rather than propagating the operators in the interaction representation, we propagate the density matrix in the Schr\"odinger representation and then 
act with the perturbation at specified times. 

For example, to simulate the $R_1$ rephasing
diagram, we first define a quantum system Hamiltonian, $H$, and transform all system operators into a basis in which $H$ is diagonal,  
$$\tilde H = e^{-S}He^{+S}$$
%\hl{where $U=e^{+S} (U^\dagger =e^{-S})$ is the unitary transformation operator which can be constructed via eigen-vectors of $H$}.
We then 
write the dipole operator (and all other operators for that matter)  in the eigenbasis as
$$
\tilde \mu = e^{-S}\mu e^{+S} = \tilde\mu_+
+\tilde \mu_- .$$

where, $\tilde\mu_+(\tilde\mu_-)$ is the part of the transformed dipole operator that leads to the excitation (de-excitation) of a ket and vice-versa for a bra.
This can be done analytically or numerically
within QuTip. Similarly, we define the initial density matrix, $\rho(-\infty)$, in the eigenbasis of $\tilde H$. It is important 
that all the operators be defined as quantum operators within the same Fock space.
QuTip will return an error if all quantum operators are not defined within the same Fock or Hilbert space.

QuTip provides an efficient 
implementation for including 
system/bath interaction using either
Redfield theory\cite{REDFIELD19651} or 
Lindblad theory\cite{Lindblad:1976aa}. Under the
Lindblad approach, the density 
matrix evolves according to 
\begin{align}
\frac{d}{dt} \rho
    = -\frac{i}{\hbar}
    [H_{sys},\rho] + 
    \sum_{i}
    \gamma_{i}
    (L_i\rho L_i^\dagger
    -\frac{1}{2}
    \left\{
    L_{i}^\dagger L_i,\rho\right\}
    )
\end{align}
where $\{L_i\}$ is a set of orthonormal operators defined in the operator space of the system Hilbert space,
%acting on the system
and $\gamma_i$ 
are real, positive rates associated with 
the bath fluctuations and dissipation. The curly brackets $\{\cdot,\cdot\}$ denote the anticommutator.

At $t=0$, we apply the first projection 
 and proceed to 
generate a specfic Liouville-space
trajectory by alternatively propagating 
and projecting. 
For example,  the $R_1$ diagram 
is specified with the QuDPy syntax as 
$$((Bu,0),(Ku,1),(Bd,2)).$$
We assume that the initial 
density matrix is stationary
up until time $t=0$, at which we apply the 
first interaction. In this case it is a 
bra-side (right-side) excitation, corresponding to 
$$\rho((Bu);0^+) = (\rho(-\infty)\tilde\mu_-).$$
This is not the full density matrix, simply one time-evolved contribution
following a sequence of steps through 
the Liouville space.  The full density matrix is a sum over all possible projections and propagations.  
Secondly, the interactions are local in 
time since we are assuming each pulse to 
be essentially a $\delta$-function in time. 
This term is propagated under the system/bath
Liouvillian operator, $\cal L$, to time $t_1$,
$$
\rho((Bu);t_1) = e^{i{\cal L}t_1}\rho((Bu);0^+) = e^{i{\cal L}t_1}(\rho(-\infty)\tilde\mu_-),
$$
at which time we act again, in this case 
on the ket-side excitation 
with $\tilde\mu_+$:
$$
\rho((Bu,Ku);t_1^+)= \tilde \mu_+ \rho((Bu);t_1)
= \tilde \mu_+\left(e^{i{\cal L}t_1}(\rho(-\infty)\tilde\mu_-)\right).
$$
Again, this is propagated to time $t_1+t_2$
at which time we act on the bra-side
with a de-excitation:
\begin{align}
\rho((Bu,Ku,Bd);(t_1+t_2)^+)&=  \left(
e^{i{\cal L}t_2}
\left(
\tilde\mu_+
e^{i{\cal L}t_1}
\left(
\rho(-\infty)\tilde\mu_-
\right)
\right)
\right)\tilde\mu_+.
\end{align}
Finally, we propagate 
for the third time interval
and act (on the ket-side) with the final 
de-excitation to produce the term 
contributing to the full polarization signal. 
For the case at hand, we have
\begin{align}
    R_1(t_1,t_2,t_3)
    = 
    \left\langle
    \tilde \mu_-
    e^{i{\cal L}t_3}
    \left(
     \left(
e^{i{\cal L}t_2}
\left(
\tilde\mu_+
e^{i{\cal L}t_1}
\left(
\rho(-\infty)\tilde\mu_-
\right)
\right)
\right)\tilde\mu_+
    \right)
    \right\rangle.
\end{align}
Similar expressions can be written 
for the other 5 response diagrams.
Note, that if ${\cal L}$ carried an
explicit time-dependency, then 
each propagation stem needs to be taken 
as a time-ordered operation. 
%This, too, is 
%straight-forward to implement within QuTip.

\begin{figure*}[ht]
    \centering
    \includegraphics[width=0.8\textwidth]{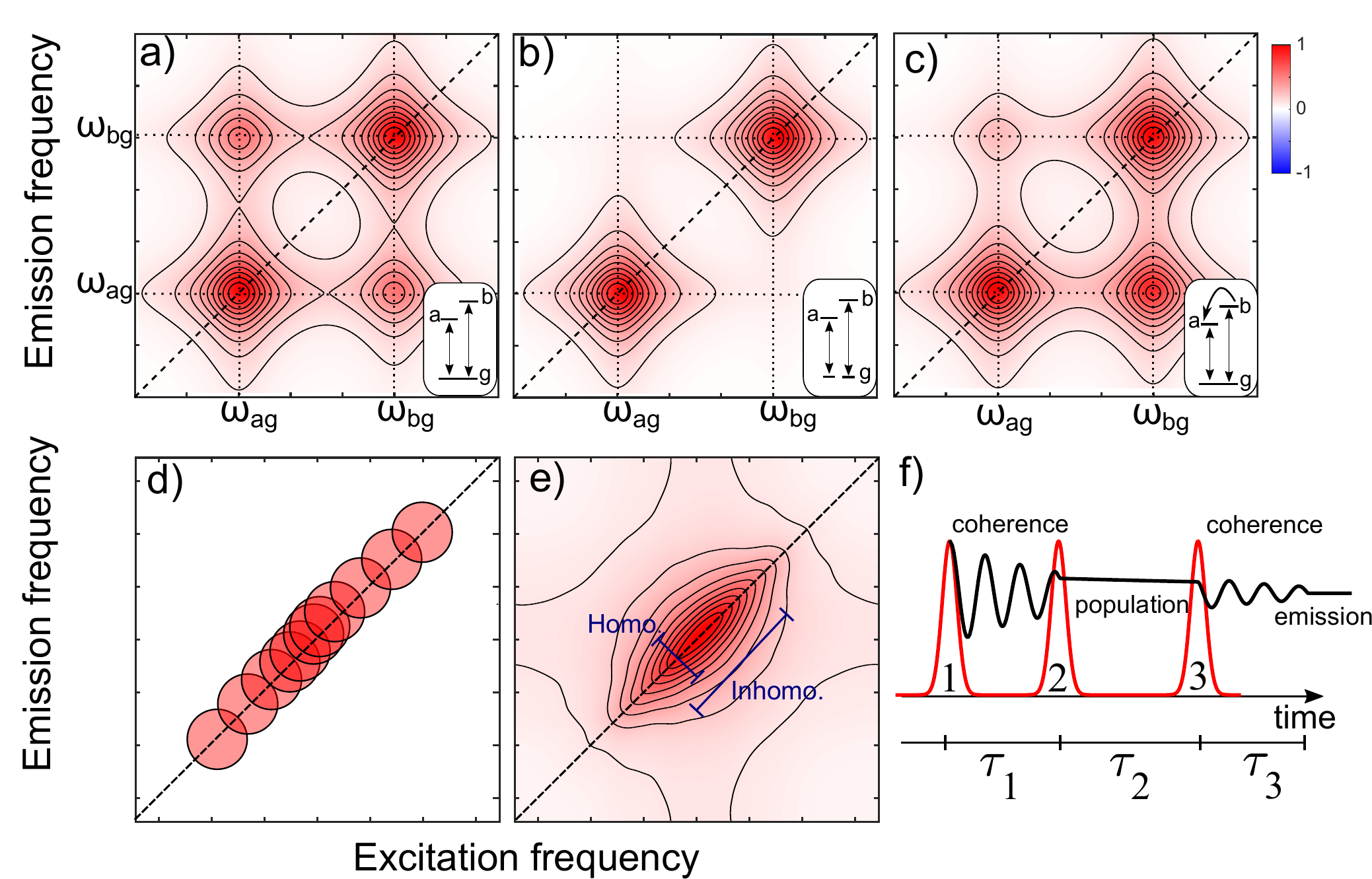}
    \caption{Example of a 2D spectra of a system with two excited levels $|a \rangle$, $|b \rangle$ and a ground state $|g \rangle$. (a) The instensity spectrum is symmetric along the diagonal since the states share a common ground, (b) the states are decoupled, (c) coupling of the $|b\rangle$ and $|a\rangle$ leads to relaxation of $|b\rangle$ towards $|a\rangle$, which translates to an asymmetric amplitude of the off-diagonal peaks. Insets in (a-c) show the energy diagrams for corresponding mechanisms and the transition frequencies are labelled by $\omega$'s. (d-e) indicate the modulus of a rephasing 2D spectrum for an inhomogeneous system that consists of a collection of emitters, where, in (d) each resonant frequency is decoupled from the others and is schematically represented by a red circle. The homogeneous width of each emitter, represented by the size of the circle, is measurable along the anti-diagonal. (e) The total signal forms an elongated peak along the diagonal (inhomogeneous broadening). (f) Pulse sequence used to perform a multi-dimensional coherence spectroscopic measurement for obtaining 3rd order polarization response.}
    \label{fig:2DExampleSpectra}
\end{figure*}

Continuing with the example of 3rd order response, in a typical experiment, one scans
the $t_1$ and $t_3$ intervals for 
a fixed $t_2$ interval. Since during $t_2$, the system 
is in a population
state of the density matrix, $t_2$ is referred to as the ``population time''. 
One then performs a 2D Fourier transform with respect to $t_1$ and $t_3$
for a fixed population time
Generally speaking, the signal along the diagonal  provides a correlation between the absorption and
emission spectra of the system for a given 
population time and the off-diagonal 
features corresponds to coherences between 
states.  Note that the off-diagonal 
coherence signal may connect both bright
and dark (non-emissive) states.
Fig.~\ref{fig:2DExampleSpectra}
gives a general overview of 
the information that
can be obtained via a 3rd order 
polarization measurement. 
In Fig.~\ref{fig:2DExampleSpectra}(a), the symmetric off-diagonal features indicate that the two states (labeled $|a\rangle$ and $|b\rangle$) share a common ground state ($|g\rangle$). In Fig.\ref{fig:2DExampleSpectra}(b),
the absence of off-diagonal features indicates that the 
states are decoupled, where as in
Fig.~\ref{fig:2DExampleSpectra}(c)
the asymmetry in the off-diagonal 
features is indicative of dynamical evolution between two excited states. 
Moreover, the overall structure
of the diagonal line-shapes gives an indication of the local environment that each state samples.  A stretching along the 
diagonal is indicative of an inhomogeneous broadening, which arises from the statistical sampling of various environments within the sample. Whereas elongation across the diagonal results from the homogeneous broadening mechanisms.
This technique
separates the 
homogeneous and inhomogeneous 
contributions to the overall line-shape.
Consequently, one obtains a
wealth of dynamical information
regarding the inner workings of 
a quantum system through the use
of coherent, multi-dimensional 
spectroscopy. However, the spectra
need to be interpreted in the 
context of a model Hamiltonian system to fully understand the spectroscopic results.

Our approach allows one to 
easily compute the quantum dynamics associated with 
an arbitrary Hamiltonian 
using QuTip (v4) quantum 
optics package.~\cite{QuTip2}
The QuTip
package is robust, easy to use, and has developed
a sizable user group spanning 
multiple areas of 
atomic and molecular physics.
We also incorporate a symbolic
approach for determining the 
irreducible/non-vanishing double-sided Feynman
diagrams ~\cite{UFSS1}, which provides a 
straight-forward way to compute the 2D responses in the time-domain.  Additionally, our method leverages the efficient Fast-Fourier Transform (FFT) capabilities 
in Numpy for efficient conversions to frequency domain.
Lastly, we note that our approach is not 
limited to 3rd order non-linear responses. 
In principle, one can use our approach to 
simulate arbitrary order experiments
with arbitrarily complicated system/bath Hamiltonians, limited only by the 
 computational resources (and time) 
 available to the user.

The disadvantage of our current implementation is that it assumes 
the interaction with the external laser field is purely impulsive, acting only at a single instant in time. However, this is not an extreme limitation 
since the experimental setups we 
are most 
interested in studying, achieve this limit 
using temporally  
narrow pulses that span the entire spectral 
frequency range of the system. Additionally, the pulse overlap is often an experimentally undesirable situation that results in coherence spikes and inclusion of additional diagrams in the signal of interest, a situation most experiments tend to avoid. \cite{UFSS1,coherenceSpike}
Finally, it is fairly straight-forward
to implement finite-sized pulses within our approach and we reserve this for a future release of our method.

\section{Downloading and 
installation}

QuDPy relies on only a few packages namely \verb|QuTip| for quantum dynamics, \verb|UFSS| for diagram generation, and supporting packages such as \verb|NumPy| and \verb|Matplotlib|. The QuDPy package can be installed with simple command
\verb|pip install qudpy==1.x.x| where \verb|1.x.x| is the version number.
The installation process installs any required packages automatically in case they are currently not installed. 
The online github repository for the package can be found here (\url{https://github.com/sa-shah/QuDPy}).

\begin{figure}
    \centering
   \includegraphics[scale=0.6]{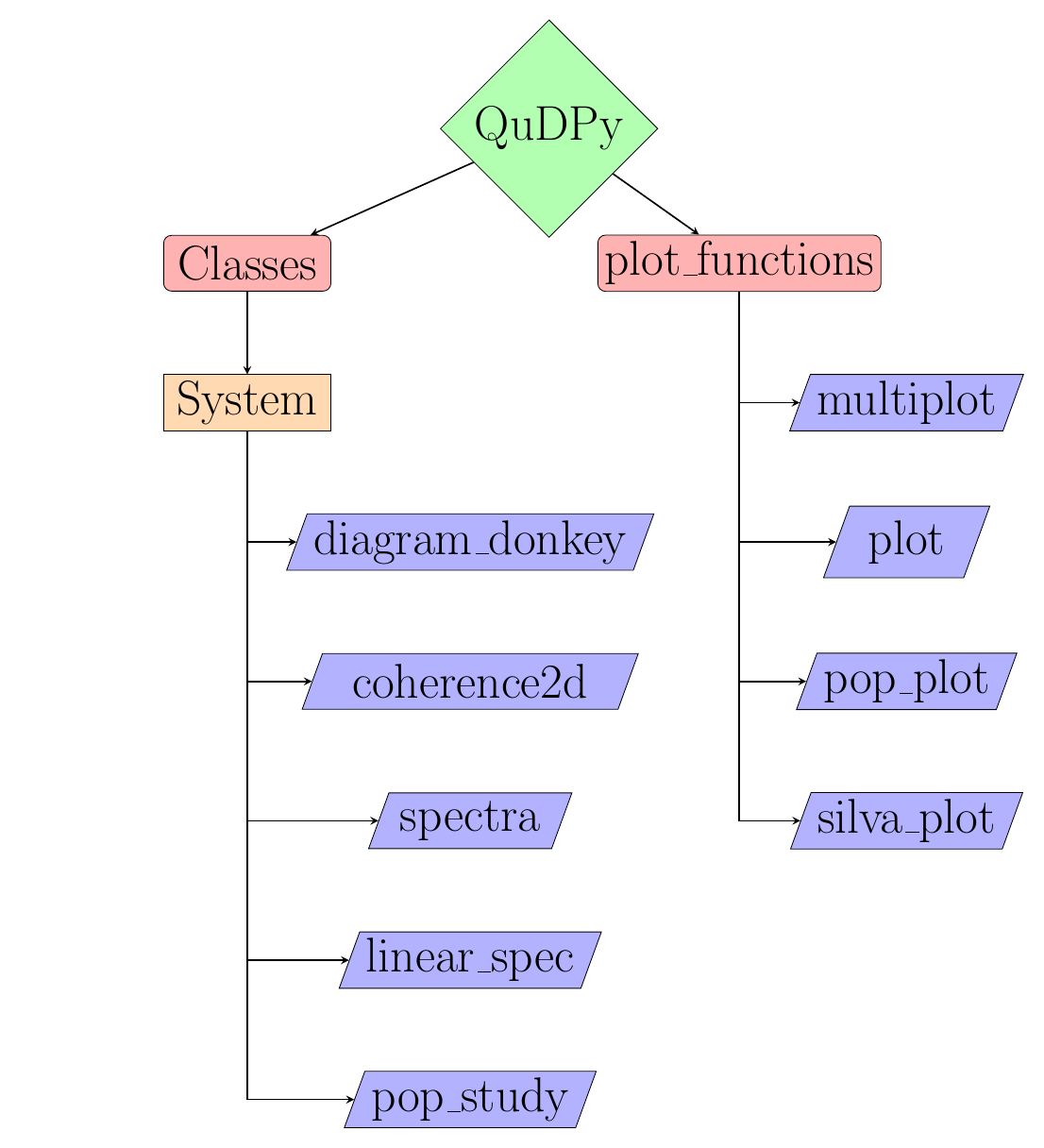}
    \caption{QuDPy structure, the subroutines are shown in red, and the functions are shown in purple.
    }
    \label{fig:flow}
\end{figure}

QuDPy itself is composed of two subroutines i.e. \verb|Classes| and \verb|plot_functions|. The structure of program is shown in Figure \ref{fig:flow}. 
The details of methods and functions available in the package (along with their input and output parameters) are provided in the Appendix.

\section{Example Calculations}
Before giving two worked examples, we give an overview of the typical workflow for using our approach. This section does not provide the complete code as it is documented in the Example 1 and 2 available on Google Colab (links available on the frontpage).
\subsection{Typical Workflow}
A typical work-flow using our approach
begins by 
first defining the system Hamiltonian
using the QuTip package and then initializing 
the diagram-generation code in UFSS using a desired phase-matching/phase-cycling condition.\cite{UFSS1,UFSS2}
The diagram-generation step is formally
agnostic of the system Hamiltonian at this point
and these two initial steps can be easily 
swapped. Furthermore, the user can provide custom diagrams without invoking UFSS, provided the diagrams are in the
format required by the UFSS routine.

The system-setup  can be accomplished
using the \verb|System()| command provided in the
\verb|QuDPy| package e.g. a custom system with Hamiltonian \verb|Hd|, light-matter interaction operator \verb|A|, dipole operator \verb|mud|, collapse operator list \verb|c_ops| and density matrix \verb|rho|, can be created by
\begin{lstlisting}
sys=System(H=Hd,
    a=A,
    u=mud,
    c_ops=[c1d,c2d,c3d,c4d],
    rho=rho)
 \end{lstlisting}
 This  is provided as utility routine for setting up
 the system and its associate quantum Fock space (that depends on the basis used to define Hamiltonian and interaction operator).
 We next provide two worked examples to illustrate 
 how to use our code for performing spectral simulations.

\subsubsection{Inclusion of light-matter interaction in quantum evolution}
In addition to the evolution under system bath interaction, the system also observes light-matter interaction in accordance with a given double-sided diagram. The diagram is generated by the Automatic Diagram Generator in UFSS for given phase-matching or phase-cycling conditions. For example, consider the case of $R^{(3)}_{1,2,3}$ in the phase matching direction $-k_1+k_2+k_3$; the corresponding diagrams can be generated by: 
\begin{lstlisting}[language=Python]
import ufss
R3rd = ufss.DiagramGenerator()
R3rd.set_phase_discrimination([(0,1),(1,0),(1,0)])
d = np.array([1]) # assigning pulse durations
R3rd.efield_times = [d, d, d, d]
# setting pulse delays
time_ordered_diagrams_rephasing = R3rd.get_diagrams([0,100,200,200]) 
[R3, R1, R2] = time_ordered_diagrams_rephasing
rephasing = [R1, R2, R3]
\end{lstlisting}
In the code above, each pulse is considered 1 fs long with pulse delays $t_1, t_2, t_3$ of $100, 200$, and $200$ fs, respectively (Note: these delays are arbitrary at this point and selected only to ensure non-overlapping pulses). The double-sided diagrams are contained in $R1, R2$, and $R3$.
These diagrams then become inputs for calculating the spectral response with appropriate system evolution. 

\subsubsection{Imposing time gating for light matter interactions}
Given a system and a double-sided diagram, the density matrix evolves with the appropriate master equation until a light-matter interaction is encountered as dictated by the diagram. At this time, the corresponding raising or lower operation is applied from the left or right on the density matrix. Afterwards, the density matrix becomes the initial density matrix for the next step of evolution under the same master equation as before till the next light-matter interaction. This patterns is repeated for all light-matter interactions. For experimentally observable spectroscopic response, only the final state after last light-matter interaction is of importance; however, states throughout the evolution can be saved for inspection.

\subsubsection{2D Coherence}
In a particularly useful experimental scheme, the coherence generated under different light-matter interactions in a double-sided diagram is utilized to probe intra-material interactions spectroscopically. In this approach, two time delays are tuned, while all others are kept fixed. The resulting final density matrices are used to generate 2D spectra by computing $\langle \mu \rangle$. The scan range for adjustable time delays is determined by either experimental feasibility or the desired spectroscopic frequency precision.

For example in $R_1$, coherences are generated during time intervals $t_1$ and $t_3$ (i.e. $|0\rangle \langle 1|$ and $|1\rangle \langle 0|$). The set of final states in corresponding 2D coherence with a $t_1$ and $t_3$ range of 200 fs each and $t_2$ fixed at 20 fs, can be generated by
\begin{lstlisting}[language=Python]
sys1 = System().
t1, t2, t3 = (200, 20, 200)
statesR1 = sys1.response2D_pop(R1, t1, t2, t3)
\end{lstlisting}
These can then be used to generate spectra with in-built functions which simply compute $\langle \mu \rangle$ from these states. 

\subsection{Fourier Transform and Resolution in Frequency Domain}
%For generating 2D coherence spectra, due to the reciprocity between the time and the frequency domain, the following observations provide valuable guidelines for efficient utilization of the computational resources and this package. The resolution in time-domain increases the scan-range of frequencies in the frequency domain; therefore, a smaller step in time domain would lead to a larger range of scanned frequencies in the spectra. \hl{The maximum difference between the eigenvalues of the system Hamiltonian can provide an estimate on the minimum time-step via Nyquist theorem for a faithful simulation.} On the other hand, a large scan-range for time will improve the resolution in the frequency domain. Consequently, to generate highly-resolved spectra, large scan-range must be selected for time-domain. An interesting situation arises when systems losses coherence (e.g. from dissipation/dephasing) in a significantly less time than that required for a desired frequency resolution in spectra. Under these conditions the additional system evolution through scan-time after the loss of coherence, is akin to zero-padding in time domain; which only results in smoothing of the spectral features in frequency domain without adding any additional information. This result can simply be achieved by using smoothing function while plotting spectra. \hl{Additionally, for inspecting the system evolution and monitoring expectation values of important operators, one can utilize a test run with} \verb|diagram_donkey| \hl{, before committing to a full-scale 2D coherence simulation.}

To generate high-resolution 2D coherence spectra, follow these guidelines for efficient utilization of computational resources:
\begin{itemize}
    \item Increase time-domain resolution for a larger scanned frequency range in the spectra.
    \item Use the maximum difference between eigenvalues of the system Hamiltonian to estimate the minimum time-step via Nyquist theorem for a faithful simulation. Note, the step size must be small enough to support QuTip differential equation solvers (QuTip will output an appropriate warning otherwise).
    \item Select a large scan-range for time-domain for improved frequency domain resolution.
    \item If the system loses coherence quickly, avoid additional system evolution as it would only result in smoothing of spectral features. Instead, use smoothing functions while plotting the spectra.
    \item For inspection of system evolution and expectation values of important observables, utilize the \verb|diagram_donkey| test run before committing to a full-scale simulation.
\end{itemize}

 \subsection{Example 1: Excitation exchange coupling between chromophores}
 
 In the example which follows, we 
consider two coupled oscillators with Hamiltonian:
\begin{align}
H/\hbar = \omega_1 a^\dagger a + \omega_2 b^\dagger b + J(a^\dagger b + b^\dagger a)
\end{align}
Physically, this corresponds to system composed of 
two chromophores that can exchange quanta via 
the exchange coupling, $J$, which can be either
positive or negative. The transition dipole operator
is the sum of the dipole operators for each oscillator,
\begin{align}
    \hat\mu  = \mu_A (a+a^\dagger) + \mu_B (b+b^\dagger).
\end{align}
In this example, we also assume each oscillator
is coupled to a thermal bath which relaxes each 
to a thermal population. 
This is a general model for a wide-range of systems 
encountered in chemical physics.

The complete Python code for this model is provided in the Example 1 Jupyter 
notebooks on Google Colab.
\begin{figure}
    \centering
 \includegraphics[width=0.5\columnwidth]{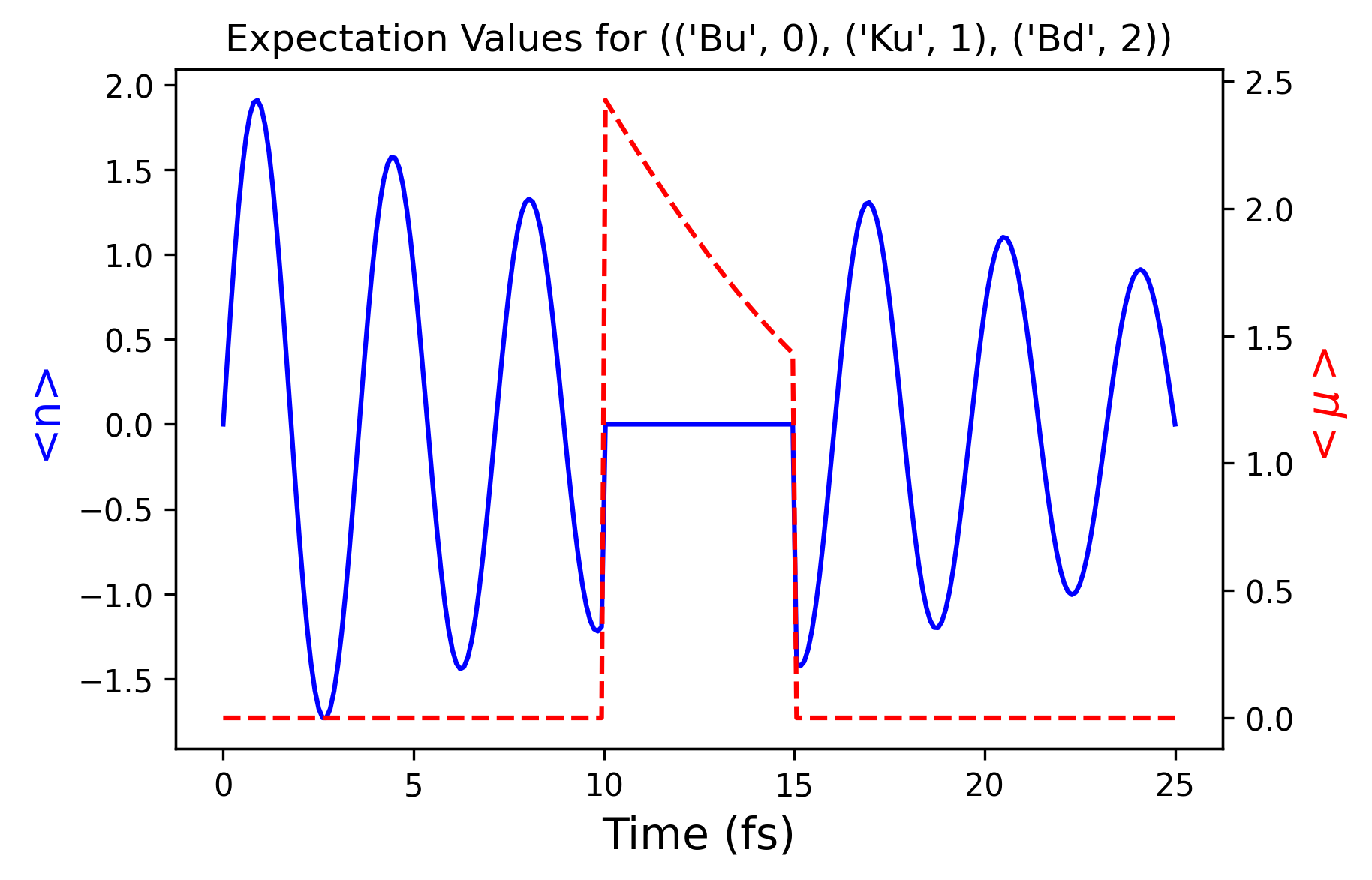}
    \caption{Results from diagram\_donkey show the expectation values of system dipole (red dashed) and excitation number (blue solid) operators for a $R_1$ diagram with a coherence time of 5 and 10 fs for the first and second coherence, and a population time of 5 fs. The results reveal the population relaxation and loss of coherence from system-bath interactions}
    \label{fig:example1}
\end{figure}
Once the system has been setup, we
are ready to compute something 
and it is useful to examine the populations 
and coherences produced by the $R_1$ diagram. 
\begin{lstlisting}[language=Python]
states = sys.diagram_donkey([0,5,10,20],[R1],r=10)
\end{lstlisting}
The pulse arrival times of $0, 5, 10, 20$ fs along with a resolution of 10 steps per fs, are selected arbitrarily to illustrate the functionality. This produces a plot showing the populations and coherences versus time as shown in Fig.~\ref{fig:example1}.   
Notice that from time $t=0$ to 5 fs, the system 
evolves as a coherence $|0\rangle\langle 1|$, 
from 5 fs to 10 fs, the system is projected onto 
a population  $|1\rangle\langle 1|$ where it undergoes
non-radiative relaxation. As noted above, this time period is referred to as the \textit{population time}.  At 10 fs, it is again 
projected onto a coherence, this time  $|1\rangle\langle 0|$ where it evolves for an additional 10fs before being projected back down to the $|0\rangle\langle 0|$ ground state. 
It is important to note that the coherences 
and populations are in the eigenbasis and not with regards to the primitive (pre-diagonalized) system.

\begin{figure}
    \centering
    \includegraphics[width=0.95\columnwidth]{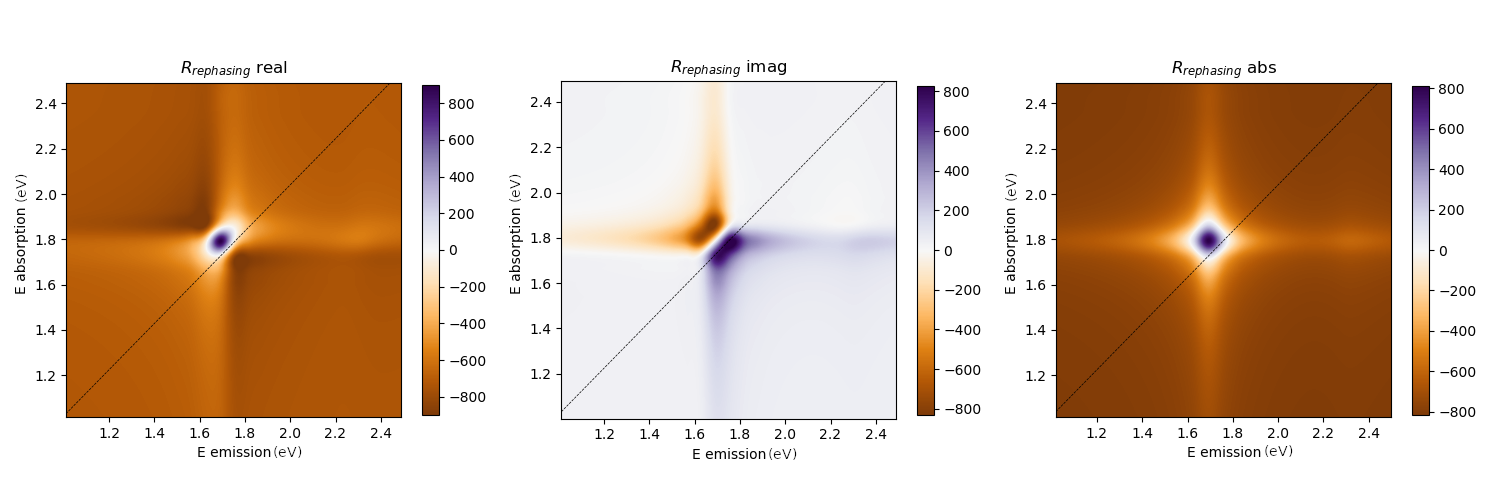}
    \includegraphics[width=0.95\columnwidth]{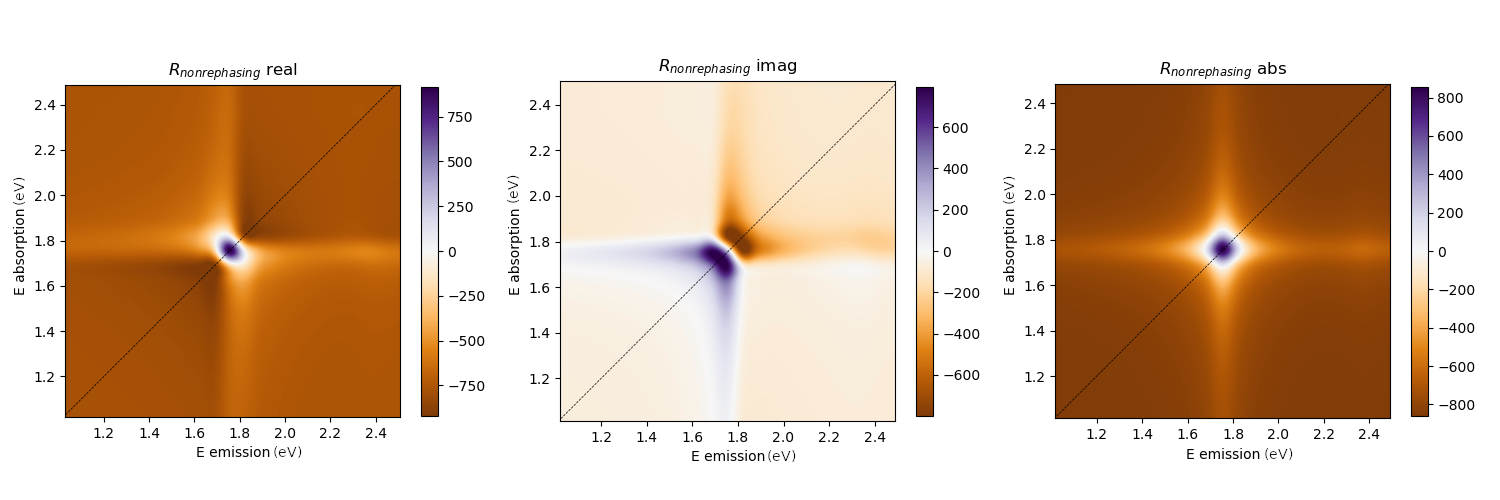}
    \caption{Frequency domain response (i.e. spectra) for two-time correlation. Top row displays the sum of rephasing diagrams $R_1$, $R_2$, and $R_3$ with left, center and right plots showing real, imaginary and absolute values, respectively. The bottom row represents the same for the sum of nonrephasing diagrams $R_4$, $R_5$ and $R_6$. These spectra only show first quadrant for the non-rephasing and fourth quandrant for rephasing (with an additional inversion of $y$-axis for rephasing). 
   % Full spectra for individual diagrams are given in appendix.
    }
    \label{fig:sample-spectra}
\end{figure}

A full simulation will involve independently scanning $t_1$ and $t_3$ for a fixed $t_2$ population time. 
This is accomplished using 
\verb|coherence2d()| routine as illustrated here.
\begin{lstlisting}[language=Python]
time_delays = [100, 5, 100]
scan_id = [0, 2]
response_list = []
states, t1, t2, dipole = sys.coherence2d(
        time_delays, diagram, scan_id)
response_list.append(np.imag(dipole))
\end{lstlisting}
Here, we are first passing a series of time-delays
$t_1$, $t_2$, and $t_3$, a list of diagrams, 
and  the indices of time delays that will be scanned e.g. 0 and 2 in this case.
Further options include the adjustment of temporal resolution for simulation and the use of parallel computing (if available); however, for clarity they are omitted from current example. The selection of scan range in this example, although arbitrary, is typical of ultrafast femtosecton experiments but it can be adjusted to the desired theoretical and practical requirements.
This step typically requires the most computational 
effort since we repeat the time propagation 
to cover the specified temporal range with 
a fine-enough grid of points to resolve the 
spectral features. The calculations return, among other things, a 2D list of dipole operator expectation values for every choice of $t_1$ and $t_3$. This response list is the main ingredient required for calculating the observable spectra.
We can now analyze the calculations and compute
the 2D spectra using the \verb|spectra()| function
as follows:
\begin{lstlisting}[language=Python]
spectra_list, extent, f1, f2 = sys.spectra(response_list)
\end{lstlisting}
Figure ~\ref{fig:sample-spectra} displays the 2D spectra for a given $t_2$ population time for this example.  The Fourier transform routine
will return the full frequency ranging spanning both positive and negative frequencies, however, in the plots only the fourth and first quadrant are shown for rephasing and nonrephasing signals respectively. Furthermore the $y$-axis is inverted for the rephasing diagrams following the conventions of ultrafast spectroscopy.

\subsection{Example 2: Cavity QED using QuDPy}

In this example we consider a driven/dissipative
open quantum system consisting of a finite number
of independent two-state atoms coupled to a quantized
mode of the radiation field (Dicke model)\cite{PhysRev.93.99}.  
The Dicke model shows a mean-field phase transition
to a super-radiant phase when 
the coupling between the light and matter crosses a critical value. The Dicke transition belongs to the Ising universality class
and has been used to model a number of  cavity quantum electrodynamics experiments.
\cite{PhysRevA.75.013804,PhysRevLett.104.130401,Baumann:2010aa}
While Dicke superradiant transition is  analogous with the lasing instability, lasing and Dicke superradiance belong to different universality classes.\cite{10.1371/journal.pone.0235197}

In this example, we shall assume that an open-quantum 
system consisting of a single cavity mode coupled to  $N$ independent 
2-level systems is probed by 
an external laser field.  
The Hamiltonian for the model reads
\begin{align}
    H = \hbar \Omega_c a^\dagger a 
    + \hbar\sum_{j=1}^N\omega_j\sigma_z(j)
    + \frac{2\lambda}{\sqrt{N}}(a+a^\dagger)\sum_j\sigma_{x}(j)
\end{align}
where $\Omega_c$ is the frequency of the (cavity) 
mode and $\{\omega_j\}$ are the transition 
frequencies of the individual spins, and $\lambda$ parameterises the coupling strength between spins and cavity mode.
In our worked model, we take 
all the spins to be identical. 
Under this assumption, we can define the
total spin operators
\begin{align}
    S_\alpha = \sum_j \sigma_{\alpha}(j)
\end{align}
which satisfies the spin algebra
$[S_x,S_y]=i\hbar S_z$.
Using these operators, the Hamiltonian 
simplifies to 
\begin{align}
    H = \hbar\Omega_c a^\dagger a
    +\hbar\omega S_z + \frac{2\lambda}{\sqrt{N}}(a+a^\dagger)S_x
\end{align}
This simplifies the numerical studies 
by reducing the Hilbert space for the 
spins from $2^N$ to $2S+1$ with $S\le N/2$.
Lastly, the third term is the coupling between the 
cavity mode and the two-level systems.  
The $\sqrt{N}$ is introduced so that the 
coupling becomes a constant in the 
limit of $N\to \infty$.
As such, we define the coupling
\begin{align}
    g = \frac{2\lambda}{\sqrt{N}}
\end{align}
and write the system Hamiltonian as 
\begin{align}
    H/\hbar = \Omega_c a^\dagger a 
    + \omega S_z
    + g (a+a^\dagger)S_x
\end{align}
In these units, 
the critical coupling occurs 
at $g_c = \sqrt{\Omega_c\omega}/2$.
The Dicke model itself is related to 
a number of other models in quantum optics. 
Specifically, when $N=1$, it is the Rabi model.  If the counter-rotating terms ($a\sigma^-(j)$ and $a^\dagger\sigma^+(j)$) are excluded, the model is termed the 
Jaynes-Cummings model for $N=1$ and the 
Tavis-Cummings model for $N>1$. 

In this example, we consider the driven/dissipative Dicke model in which the density matrix evolves according to the quantum master equation
\begin{align}
    \frac{d}{dt}\rho
    = \frac{1}{i\hbar}
    [H,\rho] 
    + \sum_\alpha
    \gamma_\alpha
    \left(
    L_\alpha \rho L^\dagger_\alpha
    -\frac{1}{2}
    \left\{L_\alpha^\dagger L_\alpha,\rho\right\}
    \right)
\end{align}
in which we include terms for
cavity decay and pumping, atomic relaxation, 
atomic dephasing, and 
collective decay. Our simulations are 
initialized by first requiring the density matrix of the system to be in a steady-state
with regards to the unperturbed quantum master equation, i.e. $d\rho_{ss}/dt=0$. 
Once $\rho_{ss} = \rho(-\infty)$ has been determined, either via numerical relaxation or 
analytically, 
the simulation proceeds as above.  For this we assume
a thermal population of 
photons in the cavity mode (as determined by the pumping intensity)
and define Lindblad operators
\begin{align}
    L_1 = \sqrt{\kappa (n_{th}+1)}a \\
     L_2 = \sqrt{\kappa n_{th}}a^\dagger
\end{align}
where $\kappa$ is the 
photon exchange rate between the cavity mode and the external bath.

\begin{figure}
    \centering
    \includegraphics[width=0.95\columnwidth]{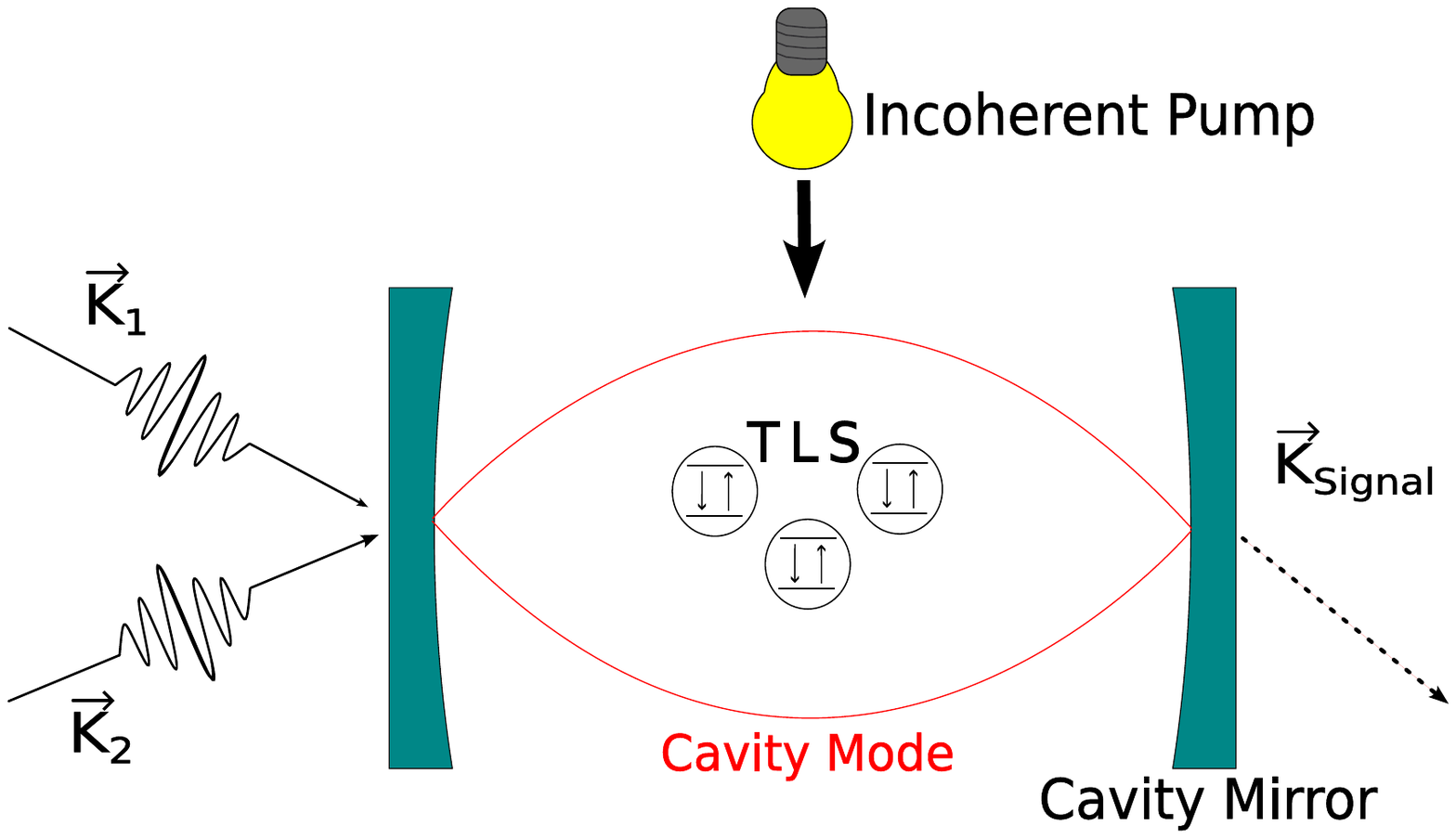}
    \caption{Dicke Model system for third order response. The cavity contains two-level systems (TLS) which interact with each other and the cavity mode (represented by red curve). 
    The cavity itself is pumped with an incoherent light source (bulb) and the whole system is probed with two pulses $K_1$ and $K_2$ . The generated signal in a proper phase-matching direction is shown as $K_{signal}$.
    }
    \label{fig:TLScavity}
\end{figure}

Within our model, we assume that the driven/dissipative cavity+spins system
is additionally probed by 
a series of perturbative ultrafast pulses
that exchange quanta with the 
cavity mode via
\begin{align}
    H_{cav/laser} \propto \mu_{cav}(b^\dagger_{laser}a + a^\dagger b_{laser}).
\end{align}
Taking the applied laser-field to be semi-classical, we can write this as
\begin{align}
    H_{cav/laser} \propto \mu_{cav}(E(t)a + E^{*}(t)a^\dagger)
\end{align}
where $E(t)$ is the electric field of the 
applied probe pulses and $\mu_{cav}$ is the transition dipole of the 
cavity.
This assumption 
is based upon the fact that
the experimental signals correspond to the macroscopic
polarizability of the entire cavity system.
Figure \ref{fig:TLScavity} provides a pictorial representation of the 
proposed experiment.

In our examples we consider the non-linear 3rd order response for a resonant
cavity $\hbar\Omega_c = \hbar\omega_j = 1$
which has a critical coupling
of $\lambda_c = \frac{\sqrt{\Omega_c\omega}}{2}=1/2$. The simulation is performed with a cavity-spin coupling strength of $0.25$, $N=6$ spins 
and 6-state basis for the cavity mode.  The cavity pumping/relaxation rate $\kappa$ is set at $0.05$. Additionally, the spin dephasing is introduced as $\sqrt{0.15} S_z$.  These are chosen 
as model parameters and are not specific to  particular physical system. 
The detailed list of parameters and additional information is provided in the relevant Jupyter notebooks in the SI/github repository. 

\begin{figure}
    \centering
    \includegraphics[width=0.95\columnwidth]{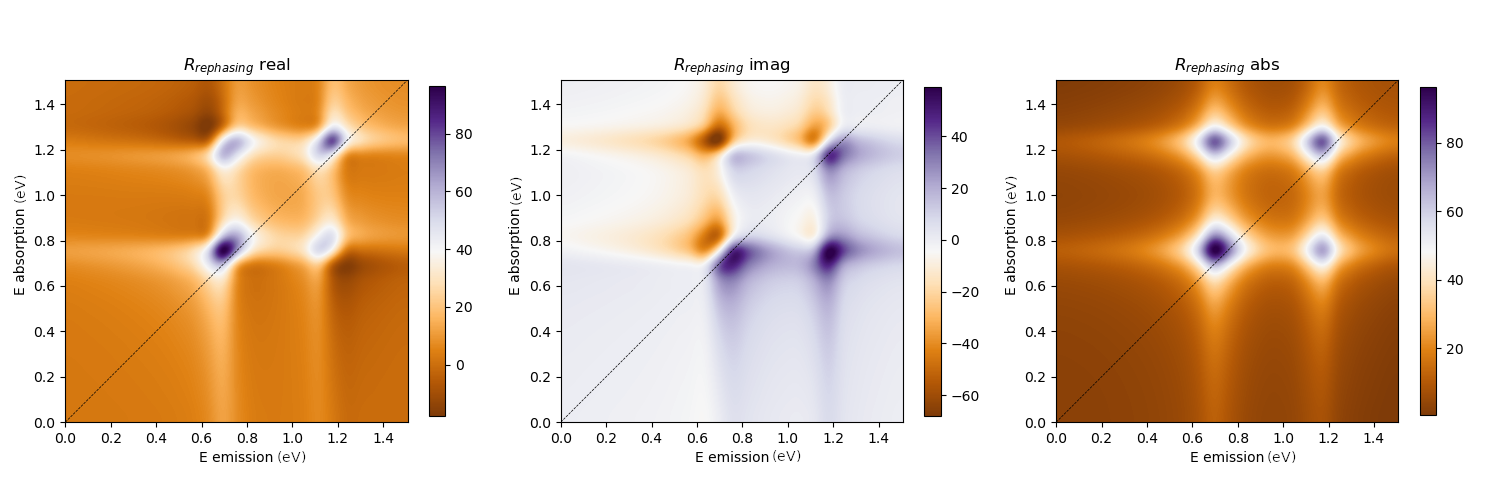}
    \includegraphics[width=0.95\columnwidth]{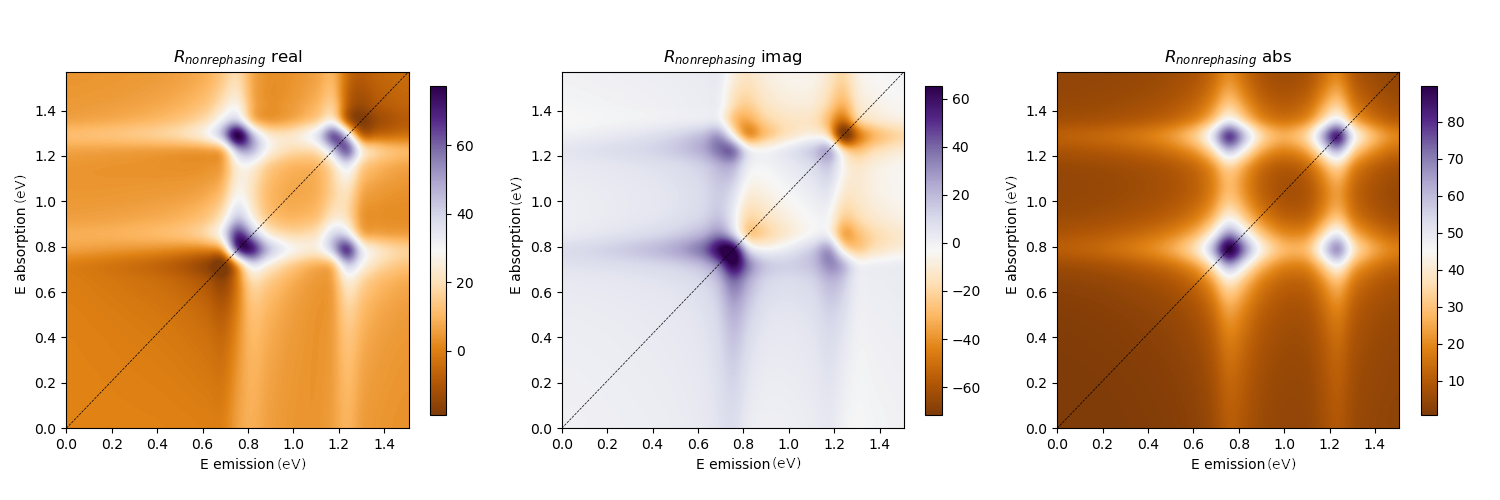}
    \caption{2D coherence spectra of response for example 2. Top row displays the sum of rephasing diagrams $R_1$, $R_2$, and $R_3$ and the bottom row represents the same for the sum of nonrephasing diagrams $R_4$, $R_5$ and $R_6$. The left, center and right columns show real, imaginary and absolute values, respectively. Only first quadrant for the non-rephasing and fourth quandrant for rephasing diagrams are shown.}
    \label{fig:sample2-spectra}
\end{figure}

Fig.~\ref{fig:sample2-spectra} shows the 
predicted 2D coherence spectra for our 
model system. In this case, one can 
clearly see symmetric off-diagonal 
coherences between the two diagonal 
peaks corresponding to the lower 
and upper polariton (LP and UP) states
of this system.  In this case, both LP and UP are connected to a common ground-state.
Physically, this can be understood 
since in setting up the problem we 
transformed the cavity operators $a$
and $a^\dagger$ to the eigenbasis of the 
Dicke Hamiltonian.  More complex dynamics 
can be introduced (for example) by 
including {\em incoherent} relaxation 
pathways via additional 
Lindblad operators. 

\section{Discussion}

We present here the first-release of 
our generalized package for simulating
non-linear spectral responses 
observed in modern ultrafast spectroscopy.
While our examples have focused upon 
optical signals, the approach we have adopted can be used to model a wide-range
of  physical problems with significant ease and control allowing the end-user
to rapidly create
model Hamiltonian systems
and compute accurate responses. 
We have avoided a detailed
analysis of the results to keep the discussion 
focused on introducing the package and its use.
Future releases of this code will 
include the facility for more complex 
pulse-shapes, time-dependent system 
Hamiltonians, electric field polarization, and the means to connect
the approach to ab-initio and molecular dynamics treatments of molecular
systems that can provide an explicit Hamiltonian with excitation manifolds and inter-band transition dipoles (as QuTip is capable of simulating the appropreate quantum dyanmics provided the explicit Hamiltonian). We believe that the code
will have tremendous utility within the ultrafast spectroscopy user community
 as well as a useful pedagogic tool for courses on modern spectroscopic methods.

\section*{Acknowledgements}

The work at Georgia Tech was funded by the National Science Foundation (DMR-1904293). 
CS acknowledges support from the School of Chemistry and Biochemistry and the College of Science at Georgia Tech. 
The work at the University of Houston was funded in part by the National Science Foundation (CHE-2102506, DMR-1903785) 
and the Robert A. Welch Foundation (E-1337).
The work at LANL was supported by the US Department of Energy through the Los Alamos National Laboratory. Los Alamos National Laboratory is operated by Triad National Security, LLC, for the National Nuclear Security Administration of U.S. Department of Energy (Contract No. 89233218CNA000001).

\section*{Author Contributions: } 

{\bf S. A. Shah}: Conceptualization, Methodology, Formal analysis, Visualization, Writing - Original draft.
{\bf Hao Li}: Formal analysis.
{\bf Eric R. Bittner}: Conceptualization, Methodology, Project administration, Writing - Review \& Editing, Funding acquisition, Supervision.
{\bf Carlos Silva}: Conceptualization,Project administration, Funding acquisition.
{\bf Andrei Piryatinski}: Conceptualization, Funding acquisition, Formal analysis.

%\bibliographystyle{elsarticle-num}
%\bibliography{References_local}

\newpage

\appendix
\section{Functions and Methods}

The default System class variables are as follows
\begin{itemize}
    \item $\hbar = 0.658211951$ in eV fs.
    \item $\omega$, system frequency with a default value of $E/\hbar$ with $E=2$ eV.
    \item $H$, Hamiltonian with default value of $\hbar \omega a^\dagger a$
    \item $a$, the system lowering operator for light-matter interaction. By default it is the lowering operator of a 3-level harmonic oscillator.
    \item $\mu$, the system dipole operator. Default value is $\mu = a^\dagger + a$
    \item $c\_ops$, a list of collapse operators for Lindblad simulation. By defaults its empty.
\end{itemize}
Furthermore, the System class can be initialized with following values.
\begin{itemize}
    \item \verb|n|, total energy levels.
    \item \verb|H|, Hamiltonian
    \item \verb|rho|, initial density matrix
    \item \verb|a|, system lowering operator encoding emission of photon.
    \item \verb|u|, system dipole operator defined as $\mu(a^\dagger+a)$. $\mu=1$ typically.
    \item \verb|c_ops|, list of collapse operators encoding system-bath coupling. 
    \item \verb|e_ops| (optional), list of operators for which expectation values are demanded in each simulation. Typically and empty list.
    \item \verb|tlist| (optional), list of time steps for simulation. Typically not required during system initialization.
    \item \verb|diagonalize|, True/False can perform Hamiltonian diagonalization and basis transformation of all operators if required.
\end{itemize}

\subsection{\textbf{Functions}}

The following functions are contained in System class  subroutine (see Figure \ref{fig:flow} for hierarchy details.
\begin{itemize}
    \item \verb|diagram_donkey|, for simulating a single trial of system %evolution through a given diagram.
    \item \verb|coherence2d|, for generating two-time coherence response.
    \item \verb|spectra|, for converting the temporal response to spectra.
    \item \verb|linear_spectra|, for calculating and generating a linear %response and spectrum from the system.
    \item \verb|pop_study|, for computing a series of 2D-coherence %response with different population 
\end{itemize}

Similarly the \verb|plot_functions| subroutine contains the following functions,
\begin{itemize}
    \item \verb|multiplot|, for plotting multiple response (temporal or frequency) simultaneously.
    \item \verb|plot|, for single spectrum or temporal response
    \item \verb|pop_plot|, for plotting a response on multiple population times.
\end{itemize}

\subsubsection{\textbf{diagram\_donkey}}
Computes and plots a single evolution of the density matrix for a list of double-sided diagrams. Mainly useful for inspection/instructional purposes.
\textbf{Inputs:}
\begin{itemize}
    \item \verb|interaction_times| (required), a list of arrival times for pulses and the last entry is time interval for detection of local oscillator.
    
    \item \verb|diagrams|(required), a list of double-sided diagrams in UFSS diagramGenerator format.
    
    \item \verb|r| (optional), temporal resolution (time steps per fs)
\end{itemize}

\textbf{Outputs:}

None, just plots the diagram.

\textbf{Notes:}

First pulse arrives at t=0.

\subsubsection{\textbf{coherence2d}}
Computes the 2D coherence plot for a single diagram with only two scan-able delays. It can be parallelized if resources are available.

\textbf{Inputs:}
\begin{itemize}
    \item \verb|time_delays| (required), list of time delays. Provide time delay for each interaction even if zero. Default is None. These should be guided by experimental and theoretical considerations for each user.
    \item \verb|diagram| (required), a double-sided diagram in UFSS diagramGenerator format. Default is None. Utilize desired phase-matching/phase-cycling conditions for diagram generation.
    \item \verb|scan_id| (required), a list of indices for the time delays in \verb|interaction_times| that have to be scanned. Default is None
    \item \verb|r| (optional), time resolution of simulation in steps per fs. Default value is 10. Should be guided by the energy eigenvalues of desired Hamiltonian.
    \item \verb|parallel| (optional), parallelization control, True or False. Default is False
\end{itemize}

\textbf{Outputs:}

\begin{itemize}
    \item A 2D list of density matrices for each pair of scanned times.
    \item A numpy array of first scan time
    \item A numpy array of second scan time
    \item A 2D list of $\langle \mu \rangle$
\end{itemize}

\textbf{Notes:}

\subsubsection{\textbf{spectra}}
Converts the list of dipoles into spectra though Fourier transform.

\textbf{Inputs:}
\begin{itemize}
    \item \verb|dipoles| (required), a list of dipoles. Default is None
    \item \verb|resolution| (optional), time resolution. Default is 10 steps per fs.
\end{itemize}

\textbf{Outputs:}
\begin{itemize}

    \item List of spectra
    \item Minimum and maximum limits of each frequency axis
    \item grid of first frequency
    \item grid of second frequency
\end{itemize}

\textbf{Notes:}
\subsubsection{\textbf{linear\_spec}}
For computing simple linear spectra from the system after any number of interaction in the start.

\textbf{Inputs:}
\begin{itemize}
    \item \verb|scan_time| (required), time interval to be simulated in fs.
    \item \verb|diagram| (optional), double-sided diagram for calculating the system response. All interactions contained in the diagram are applied at t=0. If \verb|diagram=None|, then by default a 'Bu' interaction is applied at t=0.
    \item \verb|resolution| (optional), time resolution of simulation in steps per fs.
\end{itemize}

\textbf{Outputs:}
\begin{itemize}
    \item dipole expectation value
    \item time
    \item spectrum
    \item frequency
\end{itemize}

\textbf{Notes:}

For increasing the frequency resolution, simply increase the \verb|scan_time|. For decreasing the range of frequencies, decrease the time resolution.

The function also plots the spectrum.

\subsubsection{\textbf{pop\_study}}
For calculating the nonlinear response of a double-sided diagram for a set of population times.

\textbf{Inputs:}
\begin{itemize}
    \item \verb|pop_time_list| (required), a list of population times. Default None
    \item \verb|pop_index| (required), an index of the population generating interaction. Default 1.
    \item \verb|time_delays| (required), a list of time delays between interactions. Default None
    \item \verb|diagram| (required), a double-sided diagram to be simulated. Default None
    \item \verb|scan_id| (required), list of indices of time delays to be scanned for 2D coherence plot. Default None.
    \item \verb|r| (optional), time resolution of simulation in steps per fs. Default is 10.
    \item \verb|parallel| (optional), a True/False control of parallelized computation. Default is False
\end{itemize}

\textbf{Outputs:}
\begin{itemize}
    \item list of 2D coherence response for each population time
    \item First scan time list
    \item Second scan time list
    \item List of spectra for each population time
    \item Extent of x and y-axis in spectra
    \item First frequency grid for spectra
    \item Second frequency grid for spectra.
\end{itemize}

\textbf{Notes:}

\subsubsection{\textbf{multiplot}}
Plot multiple data sets for spectral and evolution data.

\textbf{Inputs:}
\begin{itemize}
    \item \verb|data| (required), a list of spectra or dipole expectation values or any other variable of interest. Default None
    \item \verb|scan_range| (required), the min and max of both axis in the format [xmin, xmax, ymin, ymax]. Default None
    \item \verb|labels| (required), a list of label for each axis. Default None
    \item \verb|title_list| (required), a List of titles for each plot. Default None
    \item \verb|scale| (optional), scaling of the data points, two choices are 'linear' and 'log'. Default `linear'.
    \item \verb|color_map|(optional), choice of colormap. Default `viridis'.
    \item \verb|interpolation| (optional), interpolation for points in plot. Default `spline36'. Can be changed to None
    
\end{itemize}

\textbf{Outputs:} 

Does not return any output, only generates the plots.

\textbf{Notes:}

\subsubsection{\textbf{plot}}
Plots a single data set.

\textbf{Inputs:}
\begin{itemize}
    \item \verb|data| (required), a spectrum or dipole expectation value list or any other variable of interest. Default None
    \item \verb|scan_range| (required), the min and max of both axis in the format [xmin, xmax, ymin, ymax]. Default None
    \item \verb|label| (required), a list of label for each axis. Default None
    \item \verb|title_list| (required), a title for plot. Default None
    \item \verb|scale| (optional), scaling of the data points, two choices are 'linear' and 'log'. Default `linear'.
    \item \verb|color_map|(optional), choice of colormap. Default `viridis'.
    \item \verb|interpolation| (optional), interpolation for points in plot. Default `spline36'. Can be changed to None
    
\end{itemize}

\textbf{Outputs:} 

Does not return any output, only generates the plot.

\textbf{Notes:}

\subsubsection{\textbf{pop\_plot}}
Same as multiplot as of now. Will be updated in future.

\subsubsection{\textbf{silva\_plot}}
Plot multiple data sets for spectral and evolution data.

\textbf{Inputs:}
\begin{itemize}
    \item \verb|spectra_list| (required), a list of spectra or dipole expectation values or any other variable of interest. Default None
    \item \verb|scan_range| (required), the min and max of both axis in the format [xmin, xmax, ymin, ymax]. Default None
    \item \verb|labels| (required), a list of label for each axis. Default None
    \item \verb|title_list| (required), a List of titles for each plot. Default None
    \item \verb|scale| (optional), scaling of the data points, two choices are 'linear' and 'log'. Default `linear'.
    \item \verb|color_map|(optional), choice of colormap. Default `PuOr'.
    \item \verb|interpolation| (optional), interpolation for points in plot. Default `spline36'. Can be changed to None.
    \item \verb|center_scale| (optional), control for centering each data set in the list around zero by simple scale shift. default value is True.
    \item \verb|plot_sum| (optional), control for generating and plotting the total spectrum from the input list by summation of individual data sets. Default True.
    \item \verb|plot_quadrant| (optional), select a particular quadrant for the plot. Possible values are `1', `2', `3' and `4'. Default is `All'
    \item \verb|invert_y| (optional), control for inverting y-axis by changing negative values to positive. Default is False
    \item \verb|diagonals| (optional), list of True/False values for including (or excluding) the diagonal and cross diagonal reference lines. The default is [True, True]
    
\end{itemize}

\textbf{Outputs:} 

Does not return any output, only generates the plots.

\textbf{Notes:}

\end{document}